\documentclass[twocolumn,aps,showpacs,preprintnumbers,amsmath,amssymb]{revtex4-1}
\usepackage{graphicx}
\usepackage{dcolumn}
\usepackage{bm}
\usepackage{tabularx}
\usepackage{amsmath}
\usepackage{textcomp}
\usepackage{upgreek}

\begin{document}

\title{Nodal superconductivity in FeS: Evidence from quasiparticle heat transport}

\author{T. P. Ying,$^{1,\dagger}$ X. F. Lai,$^{2,\dagger}$ X. C. Hong,$^{1,\dagger}$ Y. Xu,$^1$ L. P. He,$^1$ \\J. Zhang,$^1$ M. X. Wang,$^1$ Y. J. Yu,$^1$ F. Q. Huang,$^{2,3,\ddagger}$ and S. Y. Li$^{1,4,*}$}

\affiliation{$^1$State Key Laboratory of Surface Physics, Department of Physics, and Laboratory of Advanced Materials, Fudan University, Shanghai 200433, China\\
$^2$Beijing National Laboratory for Molecular Science and State Key Laboratory of Rare Earth Materials Chemistry and Applications, College of Chemistry and Molecular Engineering, Peking University, Beijing 100871, China\\
$^3$CAS Key Laboratory of Materials for Energy Conversion and State Key Laboratory of High Performance Ceramics and Superfine Microstructure, Shanghai Institute of Ceramics, Chinese Academy of Sciences, Shanghai 200050, China\\
$^4$Collaborative Innovation Center of Advanced Microstructures, Nanjing 210093, China}

\date{\today}

\begin{abstract}

We report low-temperature heat transport measurements on superconducting iron sulfide FeS with $T_c \approx$ 5 K, which has the same crystal structure and similar electronic band structure to the superconducting iron selenide FeSe. In zero magnetic field, a significant residual linear term $\kappa_0/T$ is observed. At low field, $\kappa_0/T$ increases rapidly with the increase of field. These results provide strong evidence for nodal superconducting gap in FeS. We compare it with the sister compound FeSe, and other iron-based superconductors with nodal gap.
\end{abstract}

\maketitle

Since the discovery \cite{LaOFeAs}, the iron-based superconductors (IBSs) have been extensively studied in recent years \cite{chenxh2014review}. There are five major families of IBSs, namely, ``1111'' \cite{LaOFeAs}, ``122'' \cite{122}, ``111'' \cite{111}, ``122-selenide'' \cite{KFe2Se2}, and ``11'' \cite{11}. These families share the same FeAs(Se) layer, which resembles the CuO plane in cuprate superconductors and is responsible for the superconductivity \cite{chenxh2014review}. The electronic band structure calculation results showed strong similarities between the Fe-Se and Fe-As based superconductors, thus implied their similar superconducting nature \cite{subedi2008prb}. Furthermore, while the superconducting transition temperature ($T_c$) of bulk FeSe is only 8 K \cite{11}, it can be greatly enhanced by pressure ($T_c$ = 37 K) \cite{FeSepressure}, intercalation ($T_c$ = 46 K) \cite{yingtp2012,chenxh2015LiFeHO}, and growing single-layer FeSe on SrTiO$_3$ substrate ($T_c$ = 65 K) \cite{xueqkCPL}. In this context, the simplest IBS FeSe may provide a clean playground to investigate the superconducting pairing mechanism of IBSs.

A large step towards understanding the pairing mechanism of a superconductor is to clarify the superconducting gap structure. With the help of angle-resolved photoemission spectroscopy (ARPES), large electron pocket and nearly isotropic superconducting gap were revealed in single-layer FeSe/SrTiO$_3$ \cite{fengdl2013nm,zhouxj2013nm}, and in the intercalated (Li$_{0.8}$Fe$_{0.2}$)OHFeSe single crystal \cite{LinZhao,XHNiu}. The scanning tunneling microscopy (STM) results further indicate that single-layer FeSe/SrTiO$_3$ \cite{QFan} and potassium doped FeSe ultrathin films grown on graphitized SiC \cite{CLSong} have a plain $s$-wave pairing symmetry, with an order parameter that has the same phase on all Fermi surface sections. However, for the low-$T_c$ FeSe single crystal and FeSe single-crystalline film on the graphitized SiC(0001) substrate, the studies of superconducting gap structure gave controversial results \cite{dongjk2009,JLin,songcl2011FeSe,matsuda2014BCS}. Early thermal conductivity measurements on FeSe single crystals suggested multiple nodeless gaps \cite{dongjk2009}, which was supported by specific heat data \cite{JLin}. But the STM results on FeSe single-crystalline film and latest thermal conductivity data of stoichiometric FeSe single crystals claimed the existence of line node in the superconducting gap \cite{songcl2011FeSe,matsuda2014BCS}.

Tetragonal FeS has the same crystal structure as tetragonal FeSe, simply by replacing selenium with sulfur. Theoretically, these two compounds have very similar Fermi surface topology, with hole pockets at the Brillouin zone center and electron pockets at the zone corner \cite{subedi2008prb}. Experimentally, due to the complexity of the Fe-S phase diagram, stoichiometric tetragonal FeS was notoriously hard to obtain by high-temperature routes \cite{preFeS}. Until recently, Lai {\it et al.} applied low-temperature hydrothermal method to successfully produce stoichiometric tetragonal FeS, which shows bulk superconductivity with $T_c \approx$ 5 K \cite{lai2015FeS}. Considering their similar crystal and electronic structures, it will be very interesting to compare the superconducting properties between FeS and FeSe.

In this Rapid Communication, the thermal conductivity of well $c$-axis oriented FeS foil and FeS single crystal were measured down to 100 mK, to investigate the superconducting gap structure. Significant residual linear term $\kappa_0/T$ is observed in zero magnetic field for both samples, suggesting the existence of a nodal superconducting gap in FeS. This is further supported by the rapid field dependence of $\kappa_0/T$ at low field. The origin of this nodal superconductivity is discussed, by comparing with FeSe and other IBSs.

\begin{figure}
\includegraphics[clip,width=7cm]{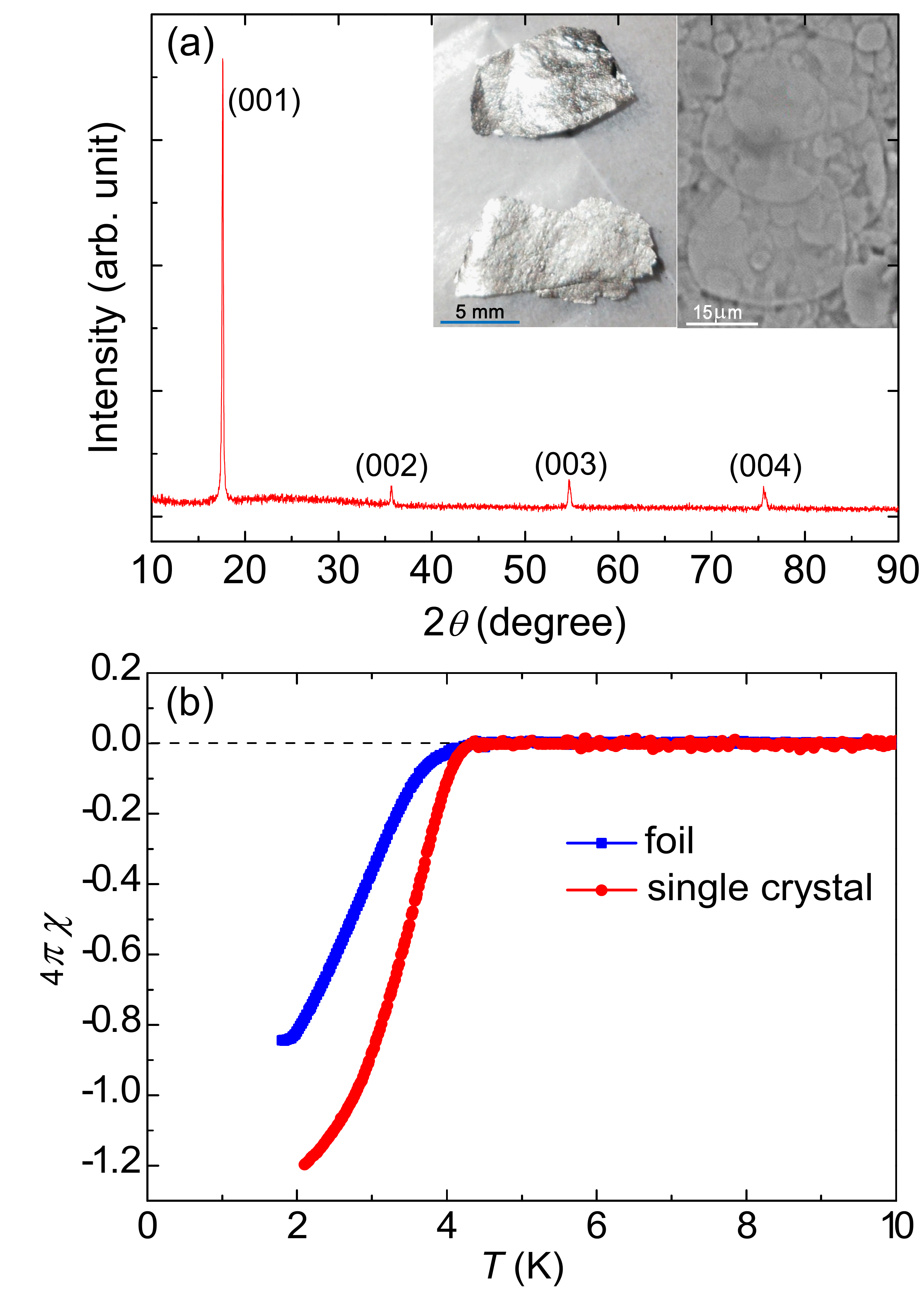}
\caption{(a) X-ray diffraction pattern of FeS foil. Only the (00$l$) Bragg peaks show up, indicating that it is well oriented along $c$ axis. Left inset: Optical image of the FeS foils. Right inset: the SEM image of the foil surface. The characterizations of FeS single crystal can be found in Ref. \cite{JZhang}. (b) The dc magnetic susceptibility at $H$ = 10 Oe for FeS foil and single crystal, measured with zero-field-cooled (ZFC) process.
}

\end{figure}

The FeS foils were obtained by the hydrothermal method described in Ref. \cite{lai2015FeS}. We also synthesized FeS single crystals by de-intercalation of K from KFe$_{2-x}$S$_2$ precursor by hydrothermal method \cite{HLin}. In this work, we present the measurements on both samples. The dc magnetization measurements were performed in a superconducting quantum interference device (SQUID) [magnetic properties measurement system (MPMS), Quantum Design]. The same samples were cut into rectangular shape and used in both resistivity and thermal conductivity measurements. The geometrical dimensions are 2.2 mm $\times$ 1.0 mm $\times$ 3.5 $\mu$m for FeS foil and 2.0 mm $\times$ 0.9 mm $\times$ 20 $\mu$m for FeS single crystal, respectively. Standard four probe method is applied with silver paint. The contacts are metallic with typical resistance of 500 m$\Omega$ for foil and 2 m$\Omega$ for single crystal at 1.5 K. The thermal conductivity was measured in a dilution refrigerator, using two RuO$_2$ chip thermometers, calibrated {\it in situ} against a reference RuO$_2$ thermometer. All the magnetic fields for resistivity and thermal conductivity measurements are applied alone the $c$ axis.

\begin{figure}
\includegraphics[clip,width=7cm]{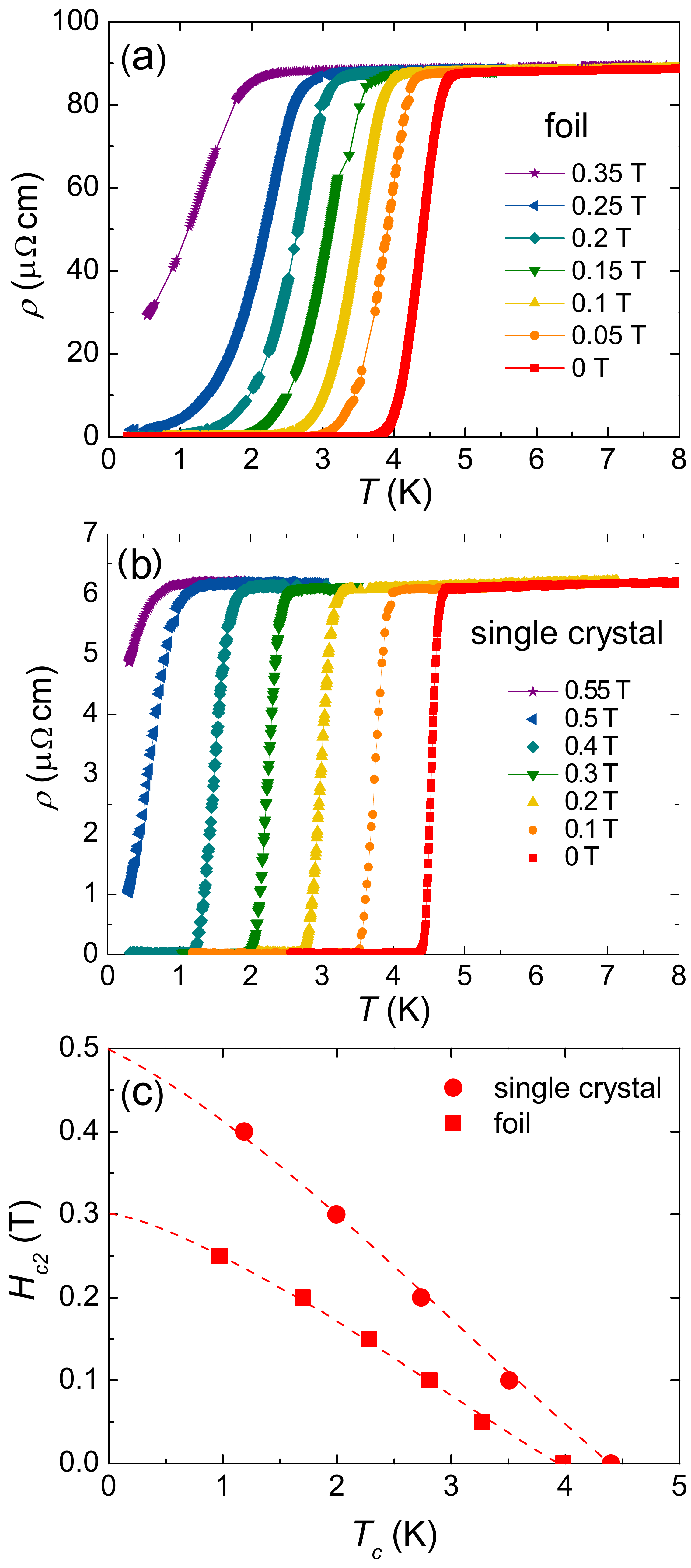}
\caption{(a) and (b) The in-plane resistivity of FeS foil and single crystal in zero and finite magnetic fields. For each field, the $T_c$ is defined at the point where $\rho$ drops to 5\% of the normal-state value. (c) The temperature dependence of upper critical field $H_{c2}$. The dashed line is a guide to the eye, which points to a roughly estimated $H_{c2}(0)^{foil} \approx$ 0.3 T and $H_{c2}(0)^{crystal} \approx$ 0.5 T.}
\end{figure}

By controlling the synthesis conditions, centimeter-size FeS foils with the thickness of several micrometers can be obtained. The FeS foil has metallic luster, as shown in the left inset of Fig. 1(a). According to the chemical analysis, no other elements were detected except for Fe and S, and the ratio Fe:S is exactly 1:1 within the experimental error. From the XRD pattern shown in Fig. 1(a), only the (00$l$) Bragg peaks show up, indicating that it is well oriented along $c$ axis. To examine whether the foil is a single crystal, we first took the SEM image on the surface. The right inset of Fig. 1(a) reveals that the surface consists of small FeS sheets. To further check it, we peeled the foil off, and the fresh surface shows similar morphology. We further performed single crystal XRD on FeS foil and found no clear diffraction spots (not shown here). Therefore, it is concluded that these FeS foils consist of well $c$-axis oriented FeS single-crystalline sheets with the size of tens micrometers, and the orientation of $a$($b$) axis for each sheet is random in the plane. According to the results of chemical analysis and XRD, there should be no impure phase in these FeS foils. Owning to the good orientation along the $c$-axis, we still measure the in-plane resistivity and thermal conductivity. The characterizations of FeS single crystal can be found in Ref. \cite{JZhang}

The dc magnetic susceptibility of FeS foil and single crystal is shown in Fig. 1(b). The diamagnetic transition of FeS foil start at 4.0 K and tends to saturate below 2 K, while that of FeS single crystal shows a slightly higher $T_c$ at about 4.3 K and a slightly larger diamagnetic signal at low temperature. Above the transition temperature, the curves are quite flat and no positive background is observed, which indicates the absence of any magnetic impurities such as Fe clusters in our samples. This is in contrast to earlier FeSe single crystals \cite{dongjk2009}, and consistent with the stoichiometric FeSe single crystals \cite{Bohmer}.

\begin{figure}
\includegraphics[clip,width=8cm]{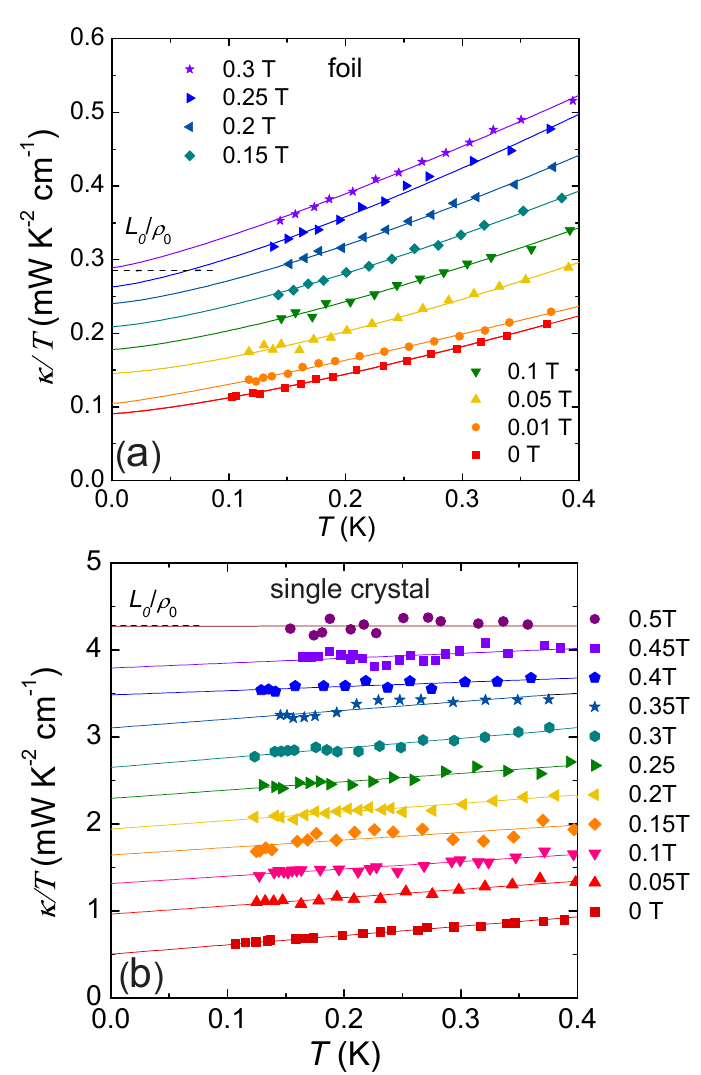}
\caption{(a) and (b) The in-plane thermal conductivity of FeS foil and single crystal in zero and finite magnetic fields. The solid lines are fits to $\kappa/T$ = $\kappa_0/T$ + $bT^{\alpha-1}$ for each curve. The $\alpha$ for FeS single crystal is fixed to 2 since its curves are roughly linear. The dashed line is the normal-state Wiedemann-Franz law expectation $L_0/\rho_0$. $L_0$ is the Lorenz number 2.45 $\times$ 10$^{-8}$ W $\Omega$ K$^{-2}$, and $\rho_0$ = 88.9 $\mu\Omega$ cm (foil) and 5.81 $\mu\Omega$ cm (single crystal).}
\end{figure}

The low-temperature in-plane resistivity in zero and finite magnetic fields is presented in Fig. 2(a) and 2(b) for FeS foil and single crystal. For each field, the $T_c$ is defined at the point where $\rho$ drops to 5\% of its normal-state value. From the zero-field $\rho$($T$) curves, FeS foil has $T_c$ = 3.98 K and FeS single crystal has $T_c$ = 4.40 K, which are consistent with the magnetization measurements. The magnetoresistance in the normal state is very weak below 0.5 T. The low-temperature resistivity from 5 to 45 K can be well fitted by the Fermi liquid behavior, $\rho(T)$ = $\rho_0$ + $AT^2$, giving $\rho_0$ = 88.9 $\mu\Omega$ cm for FeS foil and $\rho_0$ = 5.81 $\mu\Omega$ cm for FeS single crystal. As shown in Fig. 2(c), $H_{c2}^{crystal}$(0) $\approx$ 0.5 T is roughly estimated. This value is larger than that of foil, at $\approx$ 0.3 T, which is consistent with the higher $T_c$ of single crystal. To choose a slightly different $H_{c2}$(0) does not affect our discussion below. Note that the residual resistance ratio (RRR = $\rho$(298 K)/$\rho_0$) is about 6 and 40 for FeS foil and single crystal, respectively.

Ultra-low-temperature thermal conductivity measurement is an established bulk technique to probe the superconducting gap structure \cite{louis2009review}. Figure 3(a) and 3(b) plot the in-plane thermal conductivity of FeS foil and single crystal in zero and finite magnetic fields. The measured thermal conductivity contains two contributions, $\kappa$ = $\kappa_e$ + $\kappa_p$, which come from electrons and phonons, respectively. In order to separate the two contributions, all the curves below 0.4 K are fitted to $\kappa/T$ = $a$ + $bT^{\alpha-1}$ \cite{MSutherland,SYLi}, where $aT$ and $bT^{\alpha}$ represent contributions from electrons and phonons, respectively. The residual linear term $\kappa_0/T \equiv a$ is obtained by extrapolated $\kappa/T$ to $T$ = 0 K. Because of the specular reflections of phonons at the sample surfaces, the power $\alpha$ in the second term is typically between 2 and 3 \cite{MSutherland,SYLi}. We fixed $\alpha$ to 2 for FeS single crystal since its curves are roughly linear.

We first examine the Wiedemman-Franz law in the normal state of FeS. Form Fig. 3(a) and 3(b), the fittings of the data in $H_{c2}$ = 0.3 T and 0.5 T give $\kappa_{N0}/T$ = 0.283 $\pm$ 0.003 mW K$^{-2}$ cm$^{-1}$ and 4.20 $\pm$ 0.01 mW K$^{-2}$ cm$^{-1}$ for FeS foil and single crystal, respectively. Both values meet the normal-state Wiedemann-Franz law expectation $L_0/\rho_0$ = 0.276 mW K$^{-2}$ cm$^{-1}$ and 4.21 mW K$^{-2}$ cm$^{-1}$ very well. Here $L_0$ is the Lorenz number 2.45 $\times$ 10$^{-8}$ W $\Omega$ K$^{-2}$, and $\rho_0$ = 88.9 $\mu\Omega$ cm (foil) and 5.81 $\mu\Omega$ cm (single crystal). The verification of Wiedemman-Franz law in the normal state demonstrates the reliability of our thermal conductivity data.

\begin{figure}
\includegraphics[clip,width=7cm]{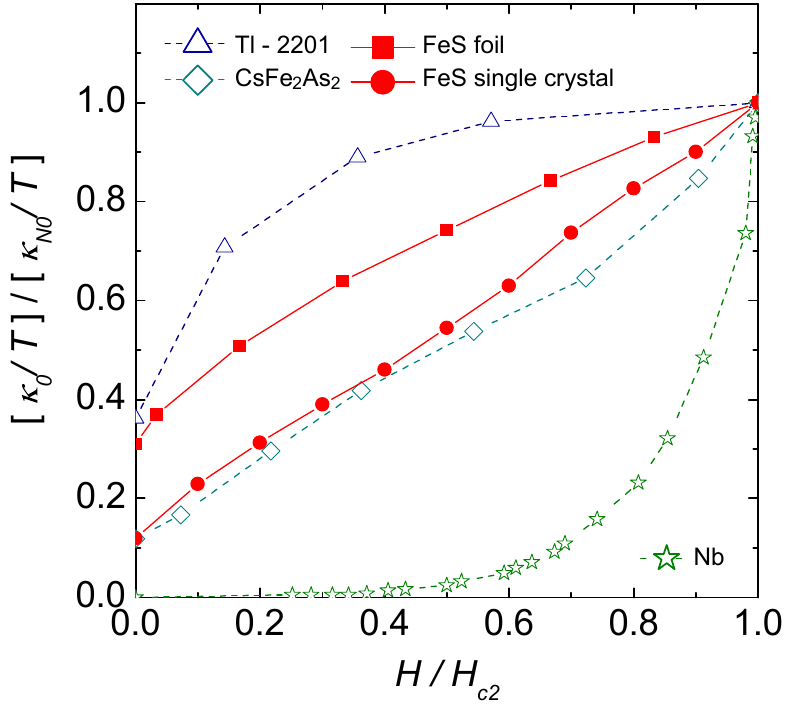}
\caption{ Normalized $\kappa_0/T$ of FeS foil and single crystal as a function of $H/H_{c2}$. Similar data of the clean single-gap $s$-wave superconductor Nb \cite{Nb}, the multiband nodal superconductor CsFe$_2$As$_2$ \cite{XCHong}, and an overdoped $d$-wave cuprate superconductor Tl-2201 \cite{Tl2201} are also shown for comparison.
}
\end{figure}

Next, in zero field, the fittings give $\kappa_0^{foil}/T$ = 92 $\pm$ 3 $\mu$W K$^{-2}$ cm$^{-1}$ and $\kappa_0^{crystal}/T$ = 503 $\pm$ 3 $\mu$W K$^{-2}$ cm$^{-1}$. Comparing with our experimental error bar $\pm$ 5 $\mu$W K$^{-2}$ cm$^{-1}$, these $\kappa_0/T$ of FeS in zero field are very significant. They are about 32\% and 12\% of their normal-state values, respectively. For $s$-wave nodeless superconductors, there are no fermionic quasiparticles to conduct heat as $T \to 0$, since all electrons become Cooper pairs \cite{louis2009review,MSutherland}. Therefore no residual linear term of $\kappa_0/T$ can be observed, as seen in V$_3$Si and NbSe$_2$ \cite{MSutherland,NbSe2}. However, for nodal superconductors, a substantial $\kappa_0/T$ in zero field contributed by the nodal quasiparticles has been found \cite{louis2009review}. For example, $\kappa_0/T$ of the overdoped $d$-wave cuprate superconductor Tl$_2$Ba$_2$CuO$_{6+\delta}$ (Tl-2201, $T_c$ = 15 K) is 1.41 mW K$^{-2}$ cm$^{-1}$, $\sim$36\% $\kappa_{N0}/T$ \cite{Tl2201}. For the $p$-wave superconductor Sr$_2$RuO$_4$ ($T_c$ = 1.5 K), $\kappa_0/T$ = 17 mW K$^{-2}$ cm$^{-1}$ was reported \cite{SrRuO4}, more than 9\% $\kappa_{N0}/T$. Therefore, the significant $\kappa_0/T$ of FeS strongly suggests that its superconducting gap has nodes.

In order to get more information on the superconducting gap structure, we then examine the field dependence of $\kappa_0/T$ \cite{louis2009review}. As shown in Fig. 3(a) and 3(b), a very small field $H$ have clearly increased the $\kappa_0/T$. After fitting all the curves and obtaining the $\kappa_0/T$ for each field, the normalized $\kappa_0/T$ as a function of $H/H_{c2}$ for FeS is plotted in Fig. 4. For comparison, data of the clean single-gap $s$-wave superconductor Nb \cite{Nb}, the multiband nodal superconductor CsFe$_2$As$_2$ \cite{XCHong}, and an overdoped $d$-wave cuprate superconductor Tl-2201 \cite{Tl2201}, are also plotted. For FeS at low field, the rapid field dependence of $\kappa_0/T$ mimics that of Tl-2201. For nodal superconductor Tl-2201, a small field can yield a quick growth in the quasiparticle density of states (DOS) due to Volovik effect, and the low field $\kappa_0(H)/T$ shows a roughly $\sqrt{H}$ dependence \cite{louis2009review}. Therefore, the rapid field dependence of $\kappa_0/T$ at low field further supports a nodal superconducting gap in FeS. However, we note that at slightly higher field, the $\kappa_0/T$ shows a slower field dependence than Tl-2201, which is not a $\sqrt{H}$ dependence. The overall field dependence of $\kappa_0/T$ for FeS is similar to that of CsFe$_2$As$_2$ and RbFe$_2$As$_2$, which were argued as multiband nodal superconductors \cite{XCHong,ZZhang}. Indeed, according to the band structure calculation of FeS, there are at least two hole pockets around the $\Gamma$ point, and two electron pockets around the $M$ point \cite{subedi2008prb}. The relatively slower field dependence of $\kappa_0/T$ indicates that the gaps in some of the Fermi surface are nodeless. For such a complex nodal $s$-wave gap structure, likely with both nodal and nodeless gaps of different magnitudes, it is hard to get a theoretical curve of $\kappa_0(H)/T$.

Having demonstrated the gap structure of FeS, we compare it with the sister compound FeSe. Early thermal conductivity and specific heat measurements on dirty FeSe single crystals suggested multiple nodeless gaps \cite{dongjk2009,JLin}. Later, for stoichiometric FeSe single-crystalline film studied by STM, the V-shaped d$I$/d$V$ and the linear dependence of the quasiparticle density of states on energy near $E_F$ explicitly reveal the existence of line nodes in the superconducting gap \cite{songcl2011FeSe}. The thermal conductivity, London penetration depth, and tunneling conductance spectrum measurements on clean stoichiometric FeSe single crystals also suggest line nodes in the gap \cite{matsuda2014BCS}. It was argued that the line nodes in FeSe are accidental, not symmetry protected \cite{matsuda2014BCS}, because the nodes are absent in disordered samples with lower RRR \cite{dongjk2009}. However, very recently, thermal conductivity, specific heat, STM, and penetration depth measurements on high-quality FeSe single crystals suggests multiple nodeless gaps again \cite{PHope,LJiao,MLi}. Since the superconducting gap structure of FeSe single crystal is still in dispute, it is not clear whether FeSe has similar nodal gap structure to FeS observed in current work.

To our knowledge, there are also some other IBSs which have nodal superconducting gap, such as LaFePO \cite{Fletcher,Hicks,MYamashida}, BaFe$_2$(As$_{1-x}$P$_x$)$_2$ \cite{YNakai,KHashimoto2,YZhang}, LiFeP \cite{Hashimoto}, BaFe$_{2-x}$Ru$_x$As$_2$ \cite{XQiu}, (K, Rb, Cs)Fe$_2$As$_2$ \cite{JKDong,Okazaki,Reid,XCHong,ZZhang}. For optimally doped BaFe$_2$(As$_{1-x}$P$_x$)$_2$, ARPES experiments found a nodal ring in the expanded $\alpha$-hole pocket at $k_z = \pi$ \cite{YZhang}. It is unknown whether this accidental nodal $s$-wave gap structure can be applied to other isovalently Ru- and P-doped iron pnictides. For KFe$_2$As$_2$, there is a hot debate on whether its nodal gap is $d$-wave or accidental nodal $s$-wave \cite{Okazaki,Reid}. A recent thermal conductivity measurements on heavily hole-doped Ba$_{1-x}$K$_x$Fe$_2$As$_2$ suggested accidental nodal $s$-wave gap \cite{XCHong2}. The finding of nodal superconducting gap in our tetragonal FeS adds another member to the list of IBSs with nodal gap.

In summary, by employing ultra-low-temperature thermal conductivity measurements, the gap structure of the newly discovered FeS superconductor is revealed. A significant residual linear term $\kappa_0/T$ is observed for both foil and single crystal samples in zero field, implying a nodal gap structure in FeS. This is further supported by our systematic studies of the evolution of $\kappa(T)/T$ with external magnetic field. More works are desired to determine the exact position of the gap nodes. Clarifying the origin of gap nodes in FeS and some other IBSs with nodal gap will give us a better understanding of their pairing mechanism.

We thank D. L. Feng, W. H. Zhang, M. Q. Ren and Y. J. Yan for valuable discussions. This work is supported by the Ministry of Science and Technology of China (National Basic Research Program No: 2012CB821402 and 2015CB921401), the Natural Science Foundation of China, Program for Professor of Special Appointment (Eastern Scholar) at Shanghai Institutions of Higher Learning, and STCSM of China (Grant No. 15XD1500200), Strategic Priority Research Program (B) of the Chinese Academy of Sciences(Grant No. XDB04040200).

Note: After we first reported the observation of nodal superconductivity in FeS foils (arXiv:1511.07717v1), a specific heat study of FeS crystals (arXiv:1512.04074) suggested nodal superconducting gap \cite{FeSwhh2}, and a theoretical study also suggested nodal $d_{x^2-y^2}$-wave superconductivity in FeS \cite{FeSwqh} . \\

$^\dagger$ These authors contribute equally to this work.\\
$^\ddagger$ huangfq@mail.sic.ac.cn\\
$^*$ shiyan$\_$li@fudan.edu.cn


\begin{thebibliography}{99}

\bibitem{LaOFeAs} Y. Kamihara, T. Watanabe, M. Hirano, and H. Hosono, Iron-based layered superconductor La(O$_{1-x}$F$_x$)FeAs (x = 0.05 $\sim$ 0.12) with T$_c$ = 26 K, J. Am. Chem. Soc. {\bf 130}, 3296 (2008).
\bibitem{chenxh2014review} X. H. Chen, P. C. Dai, D. L. Feng, T. Xiang, and F. C. Zhang, Iron-based high transition temperature superconductors, National Science Review {\bf 1}, 371 (2014).
\bibitem{122} M. Rotter, M. Tegel, and D. Johrendt, Superconductivity at 38 K in the iron arsenide Ba$_{1-x}$K$_x$Fe$_2$As$_2$, Phys. Rev. Lett. {\bf 101}, 107006 (2008).
\bibitem{111} X. Wang, Q. Liu, Y. Lv, W. Gao, L. Yang, R. Yu, F. Li, and C. Jin, The superconductivity at 18 K in LiFeAs system, Solid State Commun. {\bf 148}, 538 (2008).
\bibitem{KFe2Se2} J. Guo, S. Jin, G. Wang, S. Wang, K. Zhu, T. Zhou, M. He, and X. Chen, Superconductivity in the iron selenide K$_x$Fe$_2$Se$_2$ (0 $\leq$ x $\leq$ 1.0), Phys. Rev. B {\bf 82}, 180520 (2010).
\bibitem{11} F.-C. Hsu, J.-Y. Luo, K.-W. Yeh, T.-K. Chen, T.-W. Huang, P. M. Wu, Y.-C. Lee, Y.-L. Huang, Y.-Y. Chu, and D.-C. Yan, Superconductivity in the PbO-type structure $\alpha$-FeSe, Proc. Natl. Acad. Sci. {\bf 105}, 14262 (2008).
\bibitem{subedi2008prb} A. Subedi, L. Zhang, D. J. Singh, and M. H. Du, Density functional study of FeS, FeSe, and FeTe: Electronic structure, magnetism, phonons, and superconductivity, Phys. Rev. B {\bf 78}, 134514 (2008).
\bibitem{FeSepressure} S. Medvedev, T. M. McQueen, I. A. Troyan, T. Palasyuk, M. I. Eremets, R. J. Cava, S. Naghavi, F. Casper, V. Ksenofontov, and G. Wortmann, Electronic and magnetic phase diagram of $\beta$-Fe$_{1. 01}$Se with superconductivity at 36.7 K under pressure, Nat. Mater. {\bf 8}, 630 (2009).
\bibitem{yingtp2012} T. Ying, X. Chen, G. Wang, S. Jin, T. Zhou, X. Lai, H. Zhang, and W. Wang, Observation of superconductivity at 30 $\sim$ 46 K in A$_x$Fe$_2$Se$_2$ (A = Li, Na, Ba, Sr, Ca, Yb, and Eu), Sci. Rep. {\bf 2} (2012).
\bibitem{chenxh2015LiFeHO} X. F. Lu, N. Z. Wang, H. Wu, Y. P. Wu, D. Zhao, X. Z. Zeng, X. G. Luo, T. Wu, W. Bao, G. H. Zhang, F. Q. Huang, Q. Z. Huang and X. H. Chen, Coexistence of superconductivity and antiferromagnetism in (Li$_{0.8}$Fe$_{0.2}$)OHFeSe, Nat. Mater. {\bf 14}, 325 (2015)
\bibitem{xueqkCPL} Q. Y. Wang, Z. Li, W. H. Zhang, Z. C. Zhang, J. S. Zhang, W. Li, H. Ding, Y. B. Ou, P. Deng, K. Chang, J. Wen, C. L. Song, K. He, J. F. Jia, S. H. Ji, Y. Y. Wang, L. L. Wang, X. Chen, X. C. Ma, and Q. K. Xue, Interface-induced high-temperature superconductivity in single unit-cell FeSe films on SrTiO$_3$, Chin. Phys. Lett. {\bf 29}, 037402 (2012).
\bibitem{fengdl2013nm} S. Tan, Y. Zhang, M. Xia, Z. Ye, F. Chen, X. Xie, R. Peng, D. Xu, Q. Fan, and H. Xu, Interface-induced superconductivity and strain-dependent spin density waves in FeSe/SrTiO$_3$ thin films, Nat. Mater. {\bf 12}, 634 (2013).
\bibitem{zhouxj2013nm} S. He, J. He, W. Zhang, L. Zhao, D. Liu, X. Liu, D. Mou, Y.-B. Ou, Q.-Y. Wang, Z. Li, L. Wang, Y. Peng, Y. Liu, C. Chen, L. Yu, G. Liu, X. Dong, J. Zhang, C. Chen, Z. Xu, X. Chen, X. Ma, Q. Xue, X. J. Zhou, Phase diagram and electronic indication of high-temperature superconductivity at 65 K in single-layer FeSe films, Nat. Mater. {\bf 12}, 605 (2013).
\bibitem{LinZhao} L. Zhao, A. Liang, D. Yuan, Y. Hu, D. Liu, J. Huang, S. He, B. Shen, Y. Xu, X. Liu, L. Yu, G. Liu, H. Zhou, Y. Huang, X. Dong, F. Zhou, Z. Zhao, C. Chen, Z. Xu, and X. J. Zhou, Common electronic origin of superconductivity in (Li,Fe)OHFeSe bulk superconductor and single-layer FeSe/SrTiO$_3$ films, Nat. Commun. {\bf 7}, 10608 (2016).
\bibitem{XHNiu} X. H. Niu, R. Peng, H. C. Xu, Y. J. Yan, J. Jiang, D. F. Xu, T. L. Yu, Q. Song, Z. C. Huang, Y. X. Wang, B. P. Xie, X. F. Lu, N. Z. Wang, X. H. Chen, Z. Sun, and D. L. Feng, Surface electronic structure and isotropic superconducting gap in Li$_0.8$Fe$_0.2$OHFeSe, Phys. Rev. B {\bf 92}, 060504 (2015).
\bibitem{QFan} Q. Fan, W. H. Zhang, X. Liu, Y. J. Yan, M. Q. Ren, R. Peng, H. C. Xu, B. P. Xie, J. P. Hu, T. Zhang, and D. L. Feng, Plain s-wave superconductivity in single-layer FeSe on SrTiO$_3$ probed by scanning tunnelling microscopy, Nat. Phys. {\bf 11}, 946 (2015).
\bibitem{CLSong} C. L. Song, H.-M. Zhang, Y. Zhong, X.-P. Hu, S.-H. Ji, L. Wang, K. He, X.-C. Ma, and Q.-K. Xue, Observation of double-dome superconductivity in potassium-doped FeSe thin films, Phys. Rev. Lett. {\bf 116}, 157001 (2016).
\bibitem{dongjk2009} J. K. Dong, T. Y. Guan, S. Y. Zhou, X. Qiu, L. Ding, C. Zhang, U. Patel, Z. L. Xiao, and S. Y. Li, Multigap nodeless superconductivity in FeSe$_x$: Evidence from quasiparticle heat transport, Phys. Rev. B {\bf 80}, 024518 (2009).
\bibitem{JLin} J.-Y. Lin, Y. S. Hsieh, D. A. Chareev, A. N. Vasiliev, Y. Parsons, and H. D. Yang, Coexistence of isotropic and extended $s$-wave order parameters in FeSe as revealed by low-temperature specific heat, Phys. Rev. B {\bf 84}, 220507(R) (2011).
\bibitem{songcl2011FeSe} C.-L. Song, Y.-L. Wang, P. Cheng, Y.-P. Jiang, W. Li, T. Zhang, Z. Li, K. He, L. Wang, J.-F. Jia, L. Wang, K. He, X.-C. Ma, and Q.-K. Xue, Direct observation of nodes and twofold symmetry in FeSe superconductor, Science {\bf 332}, 1410 (2011).
\bibitem{matsuda2014BCS} S. Kasahara, T. Watashige, T. Hanaguri, Y. Kohsaka, T. Yamashita, Y. Shimoyama, Y. Mizukami, R. Endo, H. Ikeda, and K. Aoyama, Field-induced superconducting phase of FeSe in the BCS-BEC cross-over, Proc. Natl. Acad. Sci. {\bf 111}, 16309 (2014).
\bibitem{preFeS} S. D. Scott, Experimental methods in sulfide synthesis, Rev. Mineral. {\bf 1}, S1 (1974).
\bibitem{lai2015FeS} X. Lai, H. Zhang, Y. Wang, X. Wang, X. Zhang, J. Lin, and F. Huang, Observation of superconductivity in tetragonal FeS, J. Am. Chem. Soc. {\bf 137}, 10148 (2015).
\bibitem{HLin} H. Lin, Y. F. Li, Q. Deng, J. Xing, J. Liu, X. Y. Zhu, H. Yang, and H. H. Wen, Multi-band superconductivity and large anisotropy in FeS crystals. Phys. Rev. B {\bf 93}, 144505 (2016).
\bibitem{JZhang} J. Zhang, F. L. Liu, T. P. Ying, N. N. Li, Y. Xu, L. P. He, X. C. Hong, Y. J. Yu, M. X. Wang, W. G. Yang, and S. Y. Li, Observation of two distinct superconducting domes under pressure in tetragonal FeS, arXiv:1604.05254.
\bibitem{Bohmer} A. E. Bohmer, F. Hardy, F. Eilers, D. Ernst, P. Adelmann, P. Schweiss, T. Wolf, and C. Meingast, Lack of coupling between superconductivity and orthorhombic distortion in stoichiometric single-crystalline FeSe, Phys. Rev. B {\bf 87}, 180505(R) (2013).
\bibitem{louis2009review} H. Shakeripour, C. Petrovic, and L. Taillefer, Heat transport as a probe of superconducting gap structure, New J. Phys. {\bf 11}, 055065 (2009).
\bibitem{MSutherland} M. Sutherland, D.G. Hawthorn, R.W. Hill, F. Ronning, S. Wakimoto, H. Zhang, C. Proust, E. Boaknin, C. Lupien, L. Taillefer, R. Liang, D.A. Bonn, W.N. Hardy, R. Gagnon, N.E. Hussey, T. Kimura, M. Nohara, and H. Takagi, Thermal conductivity across the phase diagram of cuprates: Low-energy quasiparticles and doping dependence of the superconducting gap, Phys. Rev. B {\bf67}, 174520 (2003).
\bibitem{SYLi} S. Y. Li, J.-B. Bonnemaison, A. Payeur, P. Fournier, C. H. Wang, X. H. Chen, and L. Taillefer, Low-temperature phonon thermal conductivity of single-crystalline Nd$_2$CuO$_4$: Effects of sample size and surface roughness, Phys. Rev. B {\bf 77}, 134501 (2008).
\bibitem{NbSe2} E. Boaknin, M.A. Tanatar, J. Paglione, D. Hawthorn, F. Ronning, R.W. Hill, M. Sutherland, L. Taillefer, J. Sonier, S.M. Hayden, and J.W. Brill, Heat Conduction in the Vortex State of NbSe$_2$: Evidence for multiband superconductivity, Phys. Rev. Lett. {\bf 90}, 117003 (2003).
\bibitem{Tl2201} C. Proust, E. Boaknin, R. W. Hill, L. Taillefer, and A. P. Mackenzie, Heat transport in a strongly overdoped cuprate: Fermi liquid and a pure d-wave BCS superconductor, Phys. Rev. Lett. {\bf 89}, 147003 (2002).
\bibitem{Nb} J. Lowell and J. Sousa, Mixed-state thermal conductivity of type II superconductors, J. Low Temp. Phys. {\bf 3}, 65 (1970).
\bibitem{SrRuO4} M. Suzuki, M. A. Tanatar, N. Kikugawa, Z. Q. Mao, Y. Maeno, and T. Ishiguro, Universal heat transport in Sr$_2$RuO$_4$, Phys. Rev. Lett. {\bf 88}, 227004 (2002).
\bibitem{XCHong} X. C. Hong, X. L. Li, B. Y. Pan, L. P. He, A. F. Wang, X. G. Luo, X. H. Chen, and S. Y. Li, Nodal gap in iron-based superconductor CsFe$_2$As$_2$ probed by quasiparticle heat transport, Phys. Rev. B {\bf 87}, 144502 (2013).
\bibitem{ZZhang} Z. Zhang, A. F. Wang, X. C. Hong, J. Zhang, B. Y. Pan, J. Pan, Y. Xu, X. G. Luo, X. H. Chen, and S. Y. Li, Heat transport in RbFe$_2$As$_2$ single crystals: Evidence for nodal superconducting gap, Phys. Rev. B {\bf 91}, 024502 (2015).
\bibitem{PHope} P. Bourgeois-Hope, S. Chi, D. A. Bonn, R. Liang, W. N. Hardy, T. Wolf, C. Meingast, N. Doiron-Leyraud, and L. Taillefer, Thermal conductivity of the iron-based superconductor FeSe : Nodeless gap with strong two-band character, arXiv:1603.06917.
\bibitem{LJiao} L. Jiao, C.-L. Huang, S. R\"{o}{\ss}ler, C. Koz, U. K. R\"{o}{\ss}ler, U. Schwarz, and S. Wirth, Direct evidence for multi-gap nodeless superconductivity in FeSe, arXiv:1605.01908.
\bibitem{MLi} Meng Li, N. R. Lee-Hone, Shun Chi, Ruixing Liang, W. N. Hardy, D. A. Bonn, E. Girt, and D. M. Broun, Superfluid density and microwave conductivity of FeSe superconductor:
ultra-long-lived quasiparticles and extended $s$-wave energy gap, arXiv:1605.05407.
\bibitem{Fletcher} J. D. Fletcher, A. Serafin, L. Malone, J. G. Analytis, J.-H. Chu, A. S. Erickson, I. R. Fisher, and A. Carrington, Evidence for a nodal-line Superconducting state in LaFePO, Phys. Rev. Lett. {\bf 102}, 147001 (2009).
\bibitem{Hicks} C. W. Hicks, T. M. Lippman, M. E. Huber, J. G. Analytis, J.-H. Chu, A. S. Erickson, I. R. Fisher, and K. A. Moler, Evidence for a nodal energy gap in the iron-pnictide superconductor LaFePO from penetration depth measurements by scanning SQUID susceptometry, Phys. Rev. Lett. {\bf 103}, 127003 (2009).
\bibitem{MYamashida} M. Yamashida, N. Nakata, Y. Senshu, S. Tonegawa, K. Ikada, K. Hashimoto, H. Sugawara, T. Shibauchi, and Y. Matsuda, Thermal conductivity measurements of the energy-gap anisotropy of superconducting LaFePO at low temperatures, Phys. Rev. B {\bf 80}, 220509 (2009).
\bibitem{YNakai} Y. Nakai, T. Iye, S. Kitagawa, K. Ishida, S. Kasahara, T. Shibauchi, Y. Matsuda, and T. Terashima, $^{31}$P and $^{75}$As NMR evidence for a residual density of states at zero energy in superconducting BaFe$_{2}$(As$_{0.67}$P$_{0.33}$)$_{2}$, Phys. Rev. B {\bf 81}, 020503(R) (2010).
\bibitem{KHashimoto2} K. Hashimoto, M. Yamashita, S. Kasahara, Y. Senshu, N. Nakata, S. Tonegawa, K. Ikada, A. Serafin, A. Carrington, T. Terashima, H. Ikeda, T. Shibauchi, and Y. Matsuda, Line nodes in the energy gap of superconducting BaFe$_{2}$(As$_{1-x}$P$_{x}$)$_{2}$ single crystals as seen via penetration depth and thermal conductivity, Phys. Rev. B {\bf 81}, 220501(R) (2010).
\bibitem{YZhang} Y. Zhang, Z. R. Ye, Q. Q. Ge, F. Chen, Juan Jiang, M. Xu, B. P. Xie and D. L. Feng, Nodal superconducting-gap structure in ferropnictide superconductor BaFe$_2$(As$_{0.7}$P$_{0.3}$)$_2$, Nat. Phys. {\bf 8}, 371 (2012).
\bibitem{Hashimoto} K. Hashimoto, S. Kasahara, R. Katsumata, Y. Mizukami, M. Yamashita, H. Ikeda, T. Terashima, A. Carrington, Y. Matsuda, and T. Shibauchi, Nodal versus nodeless behaviors of the order parameters of LiFeP and LiFeAs superconductors from magnetic penetration-depth measurements, Phys. Rev. Lett. {\bf 108}, 047003 (2012).
\bibitem{XQiu} X. Qiu, S. Y. Zhou, H. Zhang, B. Y. Pan, X. C. Hong, Y. F. Dai, M. J. Eom, J. S. Kim, Z. R. Ye, Y. Zhang, D. L. Feng, and S. Y. Li, Robust nodal superconductivity induced by isovalent doping in BaFe$_{2-x}$Ru$_x$As$_2$ and BaFe$_2$(As$_{1-x}$P$_x$)$_2$, Phys. Rev. X {\bf 2}, 011010 (2012).
\bibitem{JKDong} J. K. Dong, S. Y. Zhou, T. Y. Guan, H. Zhang, Y. F. Dai, X. Qiu, X. F. Wang, Y. He, X. H. Chen, and S. Y. Li, Quantum Criticality and Nodal Superconductivity in the FeAs-Based Superconductor KFe$_2$As$_2$, Phys. Rev. Lett. {\bf 104}, 087005 (2010).
\bibitem{Okazaki} K. Okazaki, Y. Ota, Y. Kotani, W. Malaeb, Y. Ishida, T. Shimojima, T. Kiss, S. Watanabe, C.-T. Chen, K. Kihou, C. H. Lee, A. Iyo, H. Eisaki, T. Saito, H. Fukazawa, Y. Kohori, K. Hashimoto, T. Shibauchi, Y. Matsuda, H. Ikeda, H. Miyahara, R. Arita, A. Chainani, and S. Shin, Octet-line node structure of superconducting order parameter in KFe$_2$As$_2$, Science {\bf 337}, 1314 (2012).
\bibitem{Reid} J.-P. Reid, M. A. Tanatar, A. Juneau-Fecteau, R. T. Gordon, S. R. Cotret, N. Doiron-Leyraud, T. Saito, H. Fukazawa, Y. Kohori, K. Kihou, C. H. Lee, A. Iyo, H. Eisaki, R. Prozorov, and L. Taillefer, Universal heat conduction in the iron arsenide superconductor KFe$_2$As$_2$: Evidence of a $d$-wave state, Phys. Rev. Lett. {\bf 109}, 087001 (2012).
\bibitem{XCHong2} X. C. Hong, A. F. Wang, Z. Zhang, J. Pan, L. P. He, X. G. Luo, X. H. Chen, and S. Y. Li, Doping evolution of the superconducting gap Structure in heavily hole-doped Ba$_{1-x}$K$_x$Fe$_2$As$_2$: a heat transport study, Chin. Phys. Lett. {\bf 32}, 127403 (2015).
\bibitem{FeSwhh2} J. Xing, H. Lin, Y. Li, S. Li, X. Zhu, H. Yang, and H.-H. Wen, Nodal superconducting gap in $\beta$-FeS, Phys. Rev. B {\bf 93}, 104520 (2016).
\bibitem{FeSwqh} Y. Yang, W.-S. Wang, H.-Y. Liu, Y.-Y. Xiang, and Q.-H. Wang, Electronic structure and $d_{x^2 - y^2}$-wave superconductivity in FeS, Phys. Rev. B {\bf 93}, 104514 (2016).



\end{thebibliography}
\end{document}